%% file: main.tex
\documentclass{Interspeech2024}
\usepackage{multicol}
\usepackage{fancyvrb}
\usepackage{multirow}
\usepackage[dvipsnames]{xcolor}
\usepackage{expex}  

\usepackage{tabularx}

\usepackage[nopostdot,nonumberlist]{glossaries}
\setacronymstyle{long-short}  
\makeglossaries
\loadglsentries{acronyms}  
\glsdisablehyper

\interspeechcameraready

\title{Towards interfacing large language models with ASR systems using confidence measures and prompting}

\name[affiliation={1,2}]{Maryam}{Naderi}
\name[affiliation={1}]{Enno}{Hermann}
\name[affiliation={1}]{Alexandre}{Nanchen}
\name[affiliation={1}]{Sevada}{Hovsepyan}
\name[affiliation={1}]{Mathew}{Magimai.-Doss}

\address{
  $^1$Idiap Research Institute, Martigny, Switzerland\\
  $^2$UniDistance Suisse, Brig, Switzerland}
\email{\{maryam.naderi,enno.hermann,alexandre.nanchen,sevada.hovsepyan,mathew\}@idiap.ch}


\newcommand{\red}[1]{\textcolor{red}{#1}}
\newcommand{\blue}[1]{\textcolor{blue}{#1}}

\begin{document}

\maketitle

\begin{abstract}
  As \glspl{LLM} grow in parameter size and capabilities, such as interaction
  through prompting, they open up new ways of interfacing with \gls{ASR}
  systems beyond rescoring n-best lists.
  This work investigates post-hoc correction of \gls{ASR} transcripts
  with \glspl{LLM}. To avoid introducing errors into likely accurate
  transcripts, we propose a range of confidence-based filtering methods. Our
  results indicate that this can improve the performance of less competitive
  \gls{ASR} systems.
\end{abstract}

\glsresetall
\section{Introduction}

Speech perception is a complex process that relies not only on acoustic
information but also on environmental information, visual cues, context, and
other factors. Consequently, speech perception in the brain is organized in a
hierarchical and highly parallel processing network, where information on
different time scales, about different linguistic units and from different
modalities is analyzed to decipher the semantic content of
speech~\cite{rauscheckerMapsStreamsAuditory2009}. Due to the reliance on these
contextual cues during speech perception, humans can be considered ``noisy
listeners'': to successfully understand the message, humans do not need to
recognize every part of
the speech they hear. Our predictive
brain can replace the missing information based on the available contextual
information~\cite{miller1963PerceptualConsequencesLinguistic,
  shannonSpeechRecognitionPrimarily1995, gwilliams2023TopdownInformationShapesa,
  sohogluPredictiveTopDownIntegration2012}.

\Gls{ASR} systems operate in a similar way. An acoustic model first processes
the speech signal and identifies linguistic units, such as phonemes. Then, a
\gls{LM}, which encodes prior knowledge about the likelihood of different word
sequences, helps to find the most likely transcription given the potentially
noisy information from the acoustic model.

The number of parameters in \glspl{LM} and their performance on numerous
benchmarks even without task-specific fine-tuning has increased so much in
recent years that we commonly refer to them as \glspl{LLM}. Especially
instruction-tuned \glspl{LLM} offer new possibilities for down-stream
applications through their prompting mechanism~\cite{Ouyang2022-instructgpt}.
\Glspl{LLM} can also work directly with very long input contexts, obviating the
need to specifically adapt \glspl{LM} to recently observed sequences~\cite{krause2018dynamic}.

\begin{figure}[ht]
  \centering
  \includegraphics[width=\columnwidth]{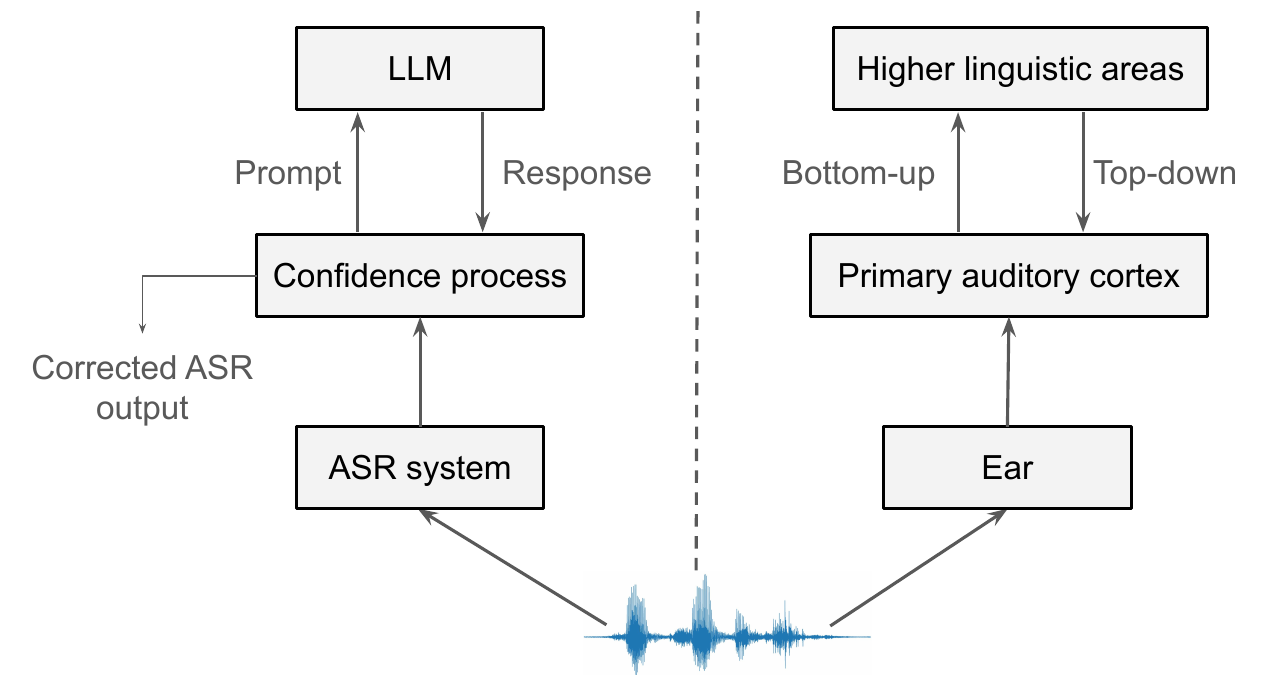}
  \caption{Proposed approach (left) and speech processing in the brain (right).}
  \label{fig:method}
\end{figure}

Motivated by these developments, we investigate combining an
\gls{ASR} system with a \gls{LLM}, where the latter is used as an
additional \gls{LM} to specifically address \gls{ASR} errors.
The primary challenge is the trade-off between leveraging LLMs to correct
errors in low-accuracy transcripts while minimizing the risk of introducing
new errors in more accurate ones. Figure~\ref{fig:method} illustrates the
proposed approach with an analogy to speech perception discussed before.
Additionally, we evaluate the impact of \gls{LLM} and \gls{ASR} model size on
the effectiveness of LLM corrections.

To reduce the chance of the \gls{LLM} introducing new errors into the
transcript, we propose three filtering methods that rely on the \gls{ASR} confidence
scores. For the first two, we let the \gls{LLM} correct only sentences where the
sentence or the lowest word confidence falls below a given threshold. For the
third method, we prompt the \gls{LLM} to only correct specific low-confidence
words.
To gain deeper insights into the LLM's behavior, we
also present concrete examples where LLM has corrected errors in the transcription and
other examples where LLM performed poorly.

The rest of this paper is organized as follows: In Section~\ref{sec:literature}
we review previous works on applying \glspl{LLM} to \gls{ASR}.
Section~\ref{sec:setup} details our experimental setup and in
Section~\ref{sec:results} we present our results and analysis.

\section{Related works}
\label{sec:literature}

With the advent of LLMs, a wide range of works have investigated how these could
improve ASR performance. Generative instruction-tuned LLMs in particular offer
new possibilities of combining ASR systems and LMs via prompting.

In hybrid speech recognition~\cite{Bourlard1994}, the decoder returns a list or
lattice of hypotheses by combining probabilities from the acoustic model and a
basic n-gram language model. These hypotheses can then be rescored with a more
powerful neural LM~\cite{Deoras2011} or LLM~\cite{udagawa2022effect}. Recent
neural end-to-end ASR approaches directly learn an LM, but can also be combined
with a separately trained one by shallow fusion or other
methods~\cite{Toshniwal2018}.

In addition to traditional integration of LMs, one can focus specifically on identifying and correcting errors in ASR outputs~\cite{Errattahi2018}. Prior works framed post-hoc ASR error correction as a spelling correction or a machine translation problem~\cite{Guo2019,Hrinchuk2020}.

In traditional n-best rescoring, only the best hypothesis is selected, although
another one could also be partially or fully correct. Chen et
al.~\cite{chen2023hyporadise} therefore instruct LLMs to generate a new
hypothesis based on all n-best options. They found that zero-shot prompting did
not yield improvements on two datasets but adapting pre-trained LLMs with
few-shot prompting, i.e. providing some example \gls{ASR} outputs and
corresponding corrections, and fine-tuning on a larger set of examples did.

Min et al.~\cite{Min2023} explored the integration of LLMs in ASR systems to
improve transcription accuracy. Their results show that directly applying the
in-context learning capabilities of the LLMs for improving ASR transcriptions
presents a significant challenge, and often leads to a higher \gls{WER}.
However, other works~\cite{Ma2023, Yang2023, radhakrishnan2023-llama} that
explored the ability of LLMs to select, rescore and correct n-best list or ASR
transcripts showed that zero and few-shot in-context learning can yield
performance gains that are comparable to rescoring by domain-tuned LMs and can
even achieve error rates below the n-best oracle level.
Other works~\cite{Li2023,He2023} have also applied \glspl{LLM} to spoken
language understanding tasks, where the focus lies on identifying the correct
intent from \gls{ASR} transcripts, rather than correcting errors.

Our work, on the contrary, focuses on giving more insights on how to effectively
use LLMs to improve ASR performance. In a few-shot, in-context learning
scenario, we evaluate the ability of LLMs to correct ASR transcripts. Similar to
Pu \textit{et al.}~\cite{Pu2023multi-stage}, we propose
filtering \gls{ASR} outputs based on confidence scores to prevent the LLM from
introducing errors into transcripts that are likely already correct. We further
analyze the errors introduced and the corrections made by LLMs. By doing so, our
work seeks to shed light on the strengths and limitations of LLMs, when applied
to ASR.

\section{Experimental setup}
\label{sec:setup}

\subsection{ASR system}

We obtain initial ASR transcriptions from Whisper~\cite{radford2023whisper}, a competitive set of models trained on 680,000~hours of transcribed speech.
We run Whisper via the \texttt{whisper-timestamped} Python
package~\cite{lintoai2023whispertimestamped}, which supports extracting sentence- and
word-level confidence scores. We compare the following models: \textit{Tiny}
(39M parameters), \textit{Medium} (769M), \textit{Large V3} (1550M).
While English-only variants exist for the smaller models, we always use the multilingual one.

\subsection{Large language model}

We use the following OpenAI ChatGPT models as \glspl{LLM} for error correction
in the \gls{ASR} transcriptions: \texttt{gpt-3.5-turbo-1106},
\texttt{gpt-3.5-turbo-0125}, and \texttt{gpt-4-0125-preview}. The GPT-3.5 turbo
models have a context window of 16,385~tokens and can understand and generate
natural language while GPT-4, with its 128,000~token context window, is a large
multimodal model (handling both text and image inputs with text outputs) and
tackles more complex problems more accurately than its
predecessors~\cite{Achiam2023GPT4TR}.
In this work, we call ChatGPT via its Python API and do not modify any default parameters.

\subsection{Confidence-based filtering}

Passing every \gls{ASR} output to the \gls{LLM} for correction risks introducing errors into correctly transcribed sentences. To mitigate this, we compare different filtering methods based on confidence scores returned by the Whisper ASR model.

Whisper internally uses tokens that are obtained with byte-pair
encoding~\cite{Sennrich2016-bpe}. Each output token is associated with a
confidence score based on its log probability. The \texttt{whisper-timestamped}
Python package~\cite{lintoai2023whispertimestamped} computes a word-level
confidence by averaging all tokens that form a word and a sentence-level
confidence by averaging all tokens in the sentence. Punctuation tokens are
excluded in either case.

We then filter the \gls{ASR} outputs that should be passed to the \gls{LLM}
based on the sentence-level or the lowest word-level confidence score in the
sentence.
We will refer to these two methods as \textit{sentence-level} and
\textit{lowest-word} confidence. For sentences above a chosen confidence
threshold, we retain the original \gls{ASR} outputs.

As a third option, we prompt the \gls{LLM} to only correct
\textit{specific words} in the \gls{ASR} transcription that fall below a certain
confidence threshold. If no words within the transcription fall below the
confidence threshold, the original \gls{ASR} transcription is retained.

\subsection{Dataset}

We evaluate our proposed approach on the English LibriSpeech corpus~\cite{panayotov2015librispeech} of audiobook recordings. We use the \texttt{dev-clean} and \texttt{dev-other} subsets for initial experiments and hyperparameter tuning and then report final results on the \texttt{test-clean} and \texttt{test-other} evaluation sets. Each of these subsets contains around 2500--3000~utterances. Speakers in the \texttt{other} portions are more challenging to recognize and lead to higher \glspl{WER}.

While these LibriSpeech subsets are not included in the Whisper training data, we cannot exclude that ChatGPT was trained on them due to the proprietary nature of the model.

\section{Results}
\label{sec:results}

In this section, we present our results in terms of \gls{WER} and \gls{CER} on identifying a suitable
prompt, comparing different \gls{ASR} models, and filtering based on confidence
scores. We also discuss and analyze the types of errors made by the
\gls{LLM}.

\subsection{Prompt selection}

We first describe our process of selecting a suitable prompt and analyze which
elements of the prompt are important for ASR performance. LLMs perform best when the prompt contains a clear description of the task. For this reason, we provided information about the task, the format of the input and the expected output, and provided two examples in the prompt.

\begin{table}[htbp]
  \caption{LLM prompts used in this work. For certain experiments, the \red{red} and/or \blue{blue} parts are added. Italic text shows examples
    provided after the system prompt, with the intended response in
    bold.}
  \label{tab:prompts}
  \scriptsize
  \centering
  \begin{tabular}{|l|}
    \hline
    You are a helpful assistant that corrects ASR errors.                               \\
    You will be presented with an ASR transcription in json format                      \\
    with key: text and your task is to correct any errors in it.                        \\
    \red{If you come across errors in ASR transcription, make corrections that}         \\
    \red{closely match the original transcription acoustically or phonetically}         \\
    \blue{If you encounter grammatical errors, provide a corrected
      version }
    \\
    \blue{adhering to proper
      grammar.}
    \\
    Provide the most probable corrected transcription in string format.                 \\
    Do not change the case, for example, lower case or upper case,                      \\
    in the transcription.                                                               \\
    Do not output any additional text that is not the corrected transcription.          \\
    Do not write any explanatory text that is not the corrected transcription.          \\
    \\
    \textit{Why not allow your silver tuff to luxuriate in a natural manner?}           \\
    \textbf{\textit{why not allow your silver tufts to luxuriate in a natural manner?}} \\
    \textit{Meanwhile, how fair did it with the flowers?}                               \\
    \textbf{\textit{Meanwhile, how fared did it with the flowers?}}                     \\
    \texttt{\textit{ASR transcription}}                                                 \\
    \hline
  \end{tabular}
\end{table}

In the prompt, we clearly explain the task of correcting \gls{ASR} errors
to the \gls{LLM}. We further describe the format of the input and expected
response and instruct the \gls{LLM} not to provide any explanatory or
additional text besides the corrected transcription. We then provide one or two example
input-output pairs for a few-shot learning scenario~\cite{chen2023hyporadise}.

We show the base prompt for our experiments with a basic description of the task
in Table~\ref{tab:prompts}. In other prompts, we explicitly instruct the
\gls{LLM} to make grammar corrections and to make changes that closely match the
input transcription acoustically or phonetically. Results for
these different prompts in Table~\ref{tab:wer_prompt} show that in particular
providing more than one example and instructing to make phonetically plausible
corrections improve the \gls{ASR} performance. For all following experiments, we
therefore use prompt~4.

\begin{table}[th]
  \caption{WER (\%, lower is better) on LibriSpeech \texttt{dev-clean} of the original Whisper \texttt{tiny} output and corrections with \texttt{gpt-3.5-turbo-1106} for different prompts.}
  \label{tab:wer_prompt}
  \centering
  \begin{tabular}{ l  r }
    \toprule
    \textbf{Prompt}                                          & \textbf{WER}  \\
    \midrule
    Original ASR output                                      & 8.51          \\
    \midrule
    1: Base prompt (with one example)                        & 7.49          \\
    2: Base prompt (with two examples)                       & 6.76          \\
    3: 2 + \blue{do correct grammar mistakes}                & 6.90          \\
    4: 2 + \red{ensure corrections are phonetically similar} & \textbf{6.65} \\
    5: 2 + \blue{3} + \red{4}                                & 6.79          \\
    \bottomrule
  \end{tabular}

\end{table}

\subsection{Influence of ASR performance}

We study the influence of \gls{ASR} performance on the \gls{LLM} corrections by
comparing Whisper models of different strength. As shown
previously~\cite{udagawa2022effect}, less competitive \gls{ASR} models ---
Whisper \textit{Tiny} and \textit{Medium} in our case --- leave more room for
improvement.

We summarize the results in terms of \gls{WER} and \gls{CER} for the original \gls{ASR}
and the \gls{LLM}-corrected transcripts (relative
change in parentheses) on both development sets in Table~\ref{tab:wer_whisper}.
While we observe improvements in WER with LLM correction in most cases,
the relative improvements in \texttt{dev-other} are smaller compared to the ones
in \texttt{dev-clean}. This suggests that correcting errors in more difficult speech data also
presents a greater challenge for the LLM model.

\begin{table}[th]
  \caption{WER and CER of the original ASR output and LLM
    corrections with \texttt{gpt-3.5-turbo-1106} for different Whisper models (relative change in parentheses).}
  \label{tab:wer_whisper}
  \centering
  \resizebox{\columnwidth}{!}{%
    \begin{tabular}{ l r r r r r r}
      \toprule
      \multicolumn{1}{c}{\textbf{Whisper}} & \multicolumn{3}{c}{\textbf{WER}} & \multicolumn{3}{c}{\textbf{CER}}                                                                       \\
      \multicolumn{1}{c}{\textbf{model}}   & ASR                              & \multicolumn{2}{c}{+LLM (rel. (\%))} & ASR     & \multicolumn{2}{c}{+LLM (rel. (\%))}                  \\
      \midrule
      \textit{dev-clean}                                                                                                                                                               \\
      Tiny                                 & 8.51                             & 6.65                                 & (-21.9) & 3.49                                 & 3.08 & (-11.7) \\
      Medium                               & 4.12                             & 3.50                                 & (-15.0) & 1.79                                 & 1.42 & (-20.7) \\
      Large V3                             & 3.11                             & 3.34                                 & (+7.4)  & 1.16                                 & 1.21 & (+4.3)  \\
      \midrule
      \textit{dev-other}                                                                                                                                                               \\
      Tiny                                 & 17.03                            & 14.87                                & (-12.7) & 8.16                                 & 7.71 & (-5.5)  \\
      Medium                               & 6.54                             & 6.19                                 & (-5.4)  & 2.96                                 & 2.90 & (-2.0)  \\
      Large V3                             & 4.62                             & 4.59                                 & (-0.6)  & 1.83                                 & 1.89 & (+3.3)  \\
      \bottomrule
    \end{tabular}}
\end{table}

\subsection{Confidence-based filtering}

\begin{figure}[!h!]
  \centering
  \includegraphics[width=\columnwidth]{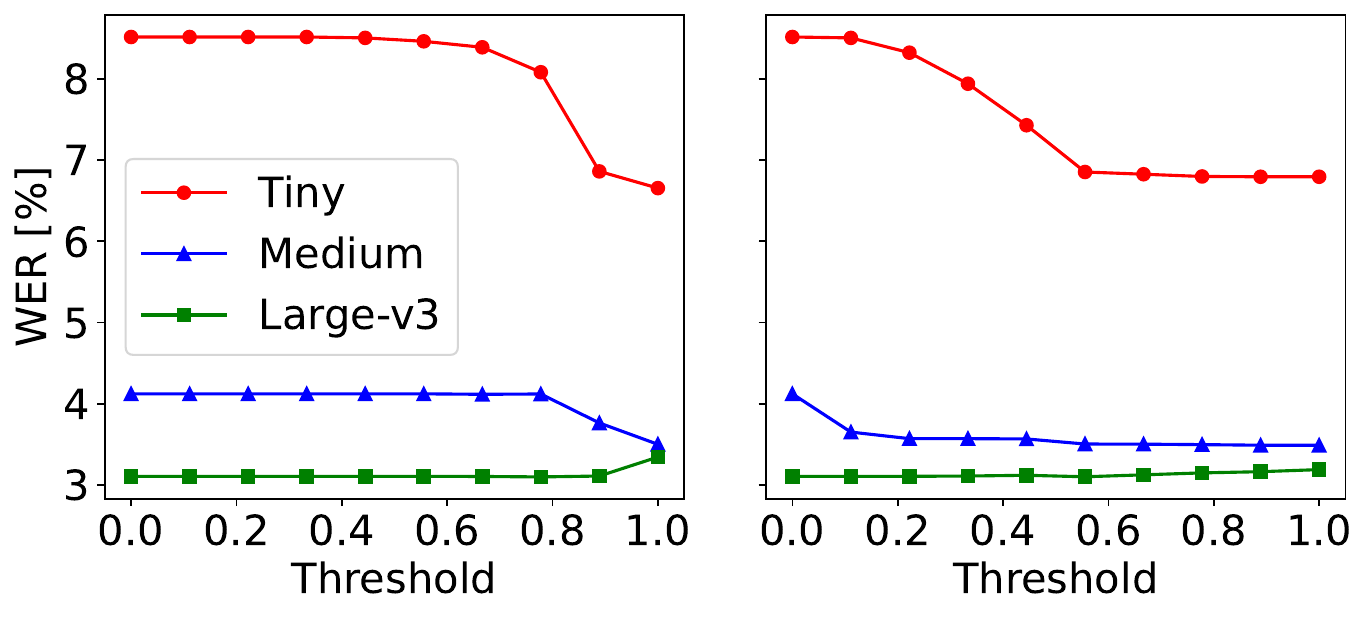}
  \caption{WER for various sentence-level (left) and lowest-word (right)
    confidence thresholds for Tiny, Medium, and
    Large~V3 Whisper models applied on \texttt{dev-clean} dataset with \texttt{gpt-3.5-turbo-1106}.}
  \label{fig:wer_threshold}
\end{figure}

In Figure~\ref{fig:wer_threshold} (left) we show the \glspl{WER} for various
\textit{sentence-level} confidence thresholds for all Whisper models on the
LibriSpeech \texttt{dev-clean} subset. Transcriptions
with a confidence score higher than the threshold are not passed to the
\gls{LLM} for correction.
The Figure shows that the optimal value for the threshold is 0.95 for \textit{Tiny} and
\textit{Medium} models while the \textit{Large} model is not sensitive to the
threshold.

Figure~\ref{fig:wer_threshold} (right) shows the effect of varying the
\textit{lowest-word} confidence thresholds. A value of~0.7 provides a good
trade-off of stable \gls{ASR} performance and reducing the number of utterances
that needs to be corrected by the \gls{LLM}.
We observed similar patterns for both methods on \texttt{dev-other}.

\subsubsection{Correction of specific words}

\begin{figure}[h!]
  \centering
  \includegraphics[width=0.9\columnwidth]{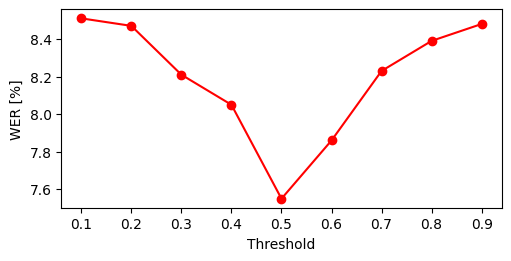}
  \caption{WER for various thresholds for specific
    low-confidence words with Tiny Whisper model applied on \texttt{dev-clean}
    dataset with \texttt{gpt-3.5-turbo-1106}.}
  \label{fig:threshold-vs-WER-certain-low}
\end{figure}

In this experiment, we pass both the ASR transcriptions and a
list of words with confidence scores below a predefined threshold to the
\gls{LLM}. \footnote{Replacing sentences 2--3 in prompt~4 from Table~\ref{tab:wer_prompt} with ``You will be presented with an ASR transcription in json format with keys: \texttt{text} and \texttt{low\_confidence\_words}, where the \texttt{text} is the ASR transcription and \texttt{low\_confidence\_words} contains the list of words in the transcription with low confidence scores.
  Your task is to correct any errors in the transcription.
  If you come across errors in ASR transcription, make sure that you correct only words from within the \texttt{low\_confidence\_words} list and your corrections should closely match the original transcription acoustically or phonetically.''}
Figure~\ref{fig:threshold-vs-WER-certain-low} presents the WER results for various
confidence thresholds.

\begin{table*}[th]
  \caption{WER on the LibriSpeech test sets of the
    original ASR output and LLM corrections with
    \texttt{gpt-3.5-turbo-0125}/\texttt{gpt-4-0125-preview} for different Whisper
    models, comparing lowest-word and sentence-level confidence. For each case, we also show what
    percentage of utterances was passed to the LLM after
    confidence-thresholding.}

  \label{tab:wer_test}
  \centering
  \begin{tabular}{ l r r r r r r r r}
    \toprule
    \textbf{Whisper}                                                     & \multicolumn{4}{c}{\textbf{test-clean}} & \multicolumn{4}{c}{\textbf{test-other}}                                  \\
    \textbf{model}                                                       & ASR
                                                                         & GPT-3.5                                 & GPT-4                                   & \% corrected  & ASR
                                                                         & GPT-3.5                                 & GPT-4                                   & \% corrected                   \\
    \midrule
    \multicolumn{5}{l}{\textit{Lowest-word confidence (threshold: 0.7)}} &                                                                                                                    \\
    Tiny                                                                 & 8.13
                                                                         & 6.55
                                                                         & 5.65                                    & 86.6\%                                  & 17.45         & 15.49
                                                                         & \textbf{13.65}                          & 94.0\%                                                                   \\
    Medium                                                               & 4.27
                                                                         &
    \textbf{3.42}                                                        & 3.54
                                                                         & 64.3\%                                  & 8.20                                    & 6.67          & 6.97  &
    72.8\%                                                                                                                                                                                    \\
    Large V3                                                             &
    \textbf{2.78}                                                        & 2.86
                                                                         & 3.21
                                                                         & 53.0\%                                  & \textbf{4.82}                           & 4.91          & 4.93
                                                                         & 60.4\%                                                                                                             \\
    \midrule
    \multicolumn{5}{l}{\textit{Sentence-level confidence (threshold: 0.95)}}                                                                                                                  \\
    Tiny                                                                 & 8.13
                                                                         & 6.56                                    & \textbf{5.63}                           & 94.5\%        &
    17.45                                                                & 15.51                                   & 13.67                                   & 98.0\%                         \\
    Medium                                                               & 4.27
                                                                         & 3.71                                    & 3.56                                    & 67.1\%        &
    8.20                                                                 & 6.77                                    & \textbf{6.62}                           & 79.8\%                         \\
    Large V3                                                             &
    \textbf{2.78}                                                        & 2.83
                                                                         & 3.13                                    & 48.4\%                                  & \textbf{4.82} & 4.93  & 4.94 &
    60.8\%                                                                                                                                                                                    \\
    \bottomrule
  \end{tabular}
\end{table*}

As the figure demonstrates, thresholds close to 0 results in a WER near the
original WER (without ChatGPT correction) but a threshold of 1 gives a higher
value for WER than the previous experiments where there was no low-confidence
word list restriction.

The WER reaches its lowest value of~7.55 at a threshold of~0.5, not matching the
performance of the sentence-level and lowest-word confidence approaches.

\subsubsection{Test set performance}
\label{sec:test}

Finally, we also present results for our selected best prompt and confidence
thresholds on the LibriSpeech \texttt{test-clean} and \texttt{test-other}
evaluation sets in Table~\ref{tab:wer_test}. The best result for each dataset
and Whisper model is highlighted in bold. Our findings indicate that, despite its higher
number of parameters, GPT-4 only outperforms GPT-3.5\footnote{Here we used
  \texttt{gpt-3.5-turbo-0125} which we found to perform similar to
  \texttt{gpt-3.5-turbo-1106} on the development sets, but is faster and more
  robust to API errors.} for Whisper
\textit{Tiny}, but does not result in additional improvements in WER for the
transcriptions of the \textit{Medium} and \textit{Large} models.

\subsection{Error analysis}
In this section, we showcase examples of Whisper \textit{Tiny} outputs on the development sets in which the LLM has corrected
errors in the transcriptions, has failed to correct or even introduced new
errors into the transcription.

\ex<ex1>
\footnotesize
\begingl
\gla REF: their fingers *** sear me like fire//
\glb ASR: their fingers see her me like fire//
\glc LLM: their fingers *** sear me like fire//
\endgl
\xe
\ex[aboveexskip=0em]<ex2>
\footnotesize
\begingl
\gla REF: damn your impertinence sir burst out burgess//
\glb ASR: dam your impertinent sur burst out burges//
\glc LLM: damn your impertinent sir burst out burgess//
\endgl
\xe
\ex[aboveexskip=0em]<ex3>
\footnotesize
\begingl
\gla REF: *** *** fedosya 's face made her anxious//
\glb ASR: the dose used to face nature *** anxious//
\glc LLM: the dose used to face nature *** anxiously//
\endgl
\xe

\noindent
Example~\getref{ex1}~and~\getref{ex2} show cases where the LLM has
corrected all or most of the errors within the ASR transcriptions of Whisper
\textit{Tiny}.  Here, \texttt{REF}, \texttt{ASR}, and \texttt{LLM} denotes
reference, ASR, and LLM-corrected transcriptions respectively.

Example~\getref{ex3} is a typical case of where the \gls{LLM}
struggles to correct the transcript because it already contains too many errors
and, for example, reconstructing proper nouns without acoustic context is challenging.
Furthermore,
Table~\ref{tab:performance} breaks down for how many utterances the \gls{LLM}
improved, worsened, or did not change the \gls{ASR} performance. We find more improvements, but also more new errors on the more challenging \texttt{dev-other} subset.

\begin{table}[th]
  \caption{Percentage of utterances where LLM improved, worsened, and did not
    change WER of Whisper Tiny outputs.}
  \label{tab:performance}
  \centering
  \begin{tabular}{ l c c c}
    \toprule
    \textbf{Dataset}   & \multicolumn{1}{c}{\textbf{Improved}} & \multicolumn{1}{c}{\textbf{Worsened}} & \multicolumn{1}{c}{\textbf{No Change}} \\
    \midrule
    \texttt{dev-clean} & 26.38                                 & 4.85                                  & 68.78                                  \\
    \texttt{dev-other} & 29.96                                 & 6.11                                  & 63.93                                  \\
    \bottomrule
  \end{tabular}
\end{table}

We also note that LLM corrections can sometimes decrease WER while
increasing CER. This occurs because any number of character changes within a
word only affects the WER by one unit ($\frac{1}{N}$ with $N$ being the number
of words in the reference transcription). However, the same changes can have a
greater impact on CER.

\ex<ex4>
\footnotesize
\begingl
\gla REF: pour mayonnaise over all chill and serve//
\glb ASR: parme a nays overall chill and serve//
\glc LLM: parmesan *** over all chill and serve//
\endgl
\xe

\noindent
Example~\getref{ex4} demonstrates this effect. The LLM
reduces WER from 57.14\% to 28.57\% in this example, while CER increases from
25.00\% to 27.50\%.

\section{Conclusions}

In this work we investigated \glspl{LLM} for \gls{ASR} error correction. Viewing \gls{ASR} systems as noisy listeners, inspired by human speech perception, we proposed filtering \gls{ASR} outputs based on confidence measures, so that the \gls{LLM} only has to focus on less accurate transcripts. Indeed, our results confirm that \glspl{LLM} especially boost \gls{ASR} performance for less competitive acoustic models because otherwise there is little room left for improvement.

We plan to investigate additional confidence estimation methods
and other \gls{ASR} systems than Whisper in future work. \Gls{LLM} outputs could also be rescored again with the acoustic model to validate if the proposed changes are acoustically plausible.
We will further consider other long-form datasets
where utterances are not evaluated one-by-one and \glspl{LLM} are expected to
provide more benefits because of their long context windows. Finally, studies
also need to be conducted on other languages, where \glspl{LLM} might not
perform as well as on English.

\section{Acknowledgements}
\ifinterspeechfinal
  This research was partially funded by the Swiss Innovation Agency Innosuisse through the flagship project IICT (grant no. PFFS-21-47) and by the Swiss National Science Foundation through the Bridge Discovery project EMIL: Emotion in the loop - a step towards a comprehensive closed-loop deep brain stimulation in Parkinson’s disease (grant no. 40B2--0$\_$194794 EMIL). We thank Olivier Bornet for providing us with a ChatGPT API key and other technical support.
\else

\fi

\bibliographystyle{IEEEtran}
\bibliography{mybib}

\end{document}

%% file: acronyms.tex
\newacronym{ASR}{ASR}{automatic speech recognition}
\newacronym{LM}{LM}{language model}
\newacronym{LLM}{LLM}{large language model}
\newacronym{POS}{POS}{part of speech}
\newacronym{WER}{WER}{word error rate}
\newacronym{CER}{CER}{character error rate}